
%
%
\input harvmac
\def\rhob{{\rho\kern-0.465em \rho}}
\def\ie{{\it i.e.}}

\def\no{\noindent}
\def\o{\over}

\def\nl{\hfill\break}
\def\alp{\alpha}\def\gam{\gamma}\def\del{\delta}

\def\sig{\sigma}
\def\ZZ{{\bf Z}}

\def\ontopss#1#2#3#4{\raise#4ex \hbox{#1}\mkern-#3mu {#2}}

\setbox\strutbox=\hbox{\vrule height12pt depth5pt width0pt}

\def\strut{\relax\ifmmode\copy\strutbox\else\unhcopy\strutbox\fi}


\nref\rKKMM{R.~Kedem, T.R.~Klassen, B.M.~McCoy and E.~Melzer,
 Stony Brook/Rutgers preprint ITP-SB-92-64/RU-92-51, hep-th/9211102,
 Phys. Lett. B (in press).}
\nref\rLepPrim{J.~Lepowsky and M.~Primc, {\it Structure of the
 standard modules for the affine Lie algebra $A_1^{(1)}$},
 Contemporary Mathematics, Vol.~46 (AMS, Providence, 1985).}
\nref\rKedMc{R.~Kedem and B.M.~McCoy, J. Stat. Phys.~(in press).}
\nref\rFNO{B.L. Feigin, T. Nakanishi and H. Ooguri, Int. J. Mod.
Phys. A7, Suppl. 1A (1992) 217}
\nref\rNRT{W.~Nahm, A.~Recknagel and M.~Terhoeven, Bonn preprint,
   hep-th/9211034.}
\nref\rTerh{M.~Terhoeven, Bonn preprint BONN-HE-92-36, hep-th/9111120.}
\nref\rKNS{A. Kuniba, T. Nakanishi, J. Suzuki, Harvard preprint HUPT-92/A069}
\nref\rDKMM{S. Dasmahapatra, R. Kedem, B.M. McCoy and E. Melzer
 (in preparation).}
\nref\rorbif{P. Fendley and P. Ginsparg, Nucl. Phys. B324 (1989) 549;\nl
 Ph. Roche, Commun. Math. Phys. 127 (1990) 395.}
\nref\rCar{J.L.~Cardy, Nucl.~Phys.~B270 (1986) 186.}
\nref\rModInv{A.~Cappelli, C.~Itzykson and J.-B.~Zuber, Nucl.~Phys.~B280 (1987)
 445; \nl D.~Gepner, Nucl.~Phys.~B287 (1987) 111.}
\nref\rRoCa{A.~Rocha-Caridi, in: {\it Vertex Operators in Mathematics
 and Physics}, ed. J.~Lepowsky {\it et al} (Springer, Berlin, 1985).}
\nref\rFFF{B.L Feigin and D.B. Fuchs, Funct. Anal. Appl. 17 (1983) 241;\nl
 G. Felder, Nucl. Phys. B317 (1989) 215.}
\nref\randrews{G.E. Andrews, {\it The Theory of Partitions}
  (Addison-Wesley, London, 1976).}
\nref\rSlater{L.J. Slater, Proc. London Math. Soc. 54 (1953) 147.}
\nref\rStan{R.P. Stanley, {\it Ordered Structures and Partitions},
 Mem. Amer. Math. Soc. 119 (1972).}
\nref\rJKM{J.D. Johnson, S. Krinsky and B.M. McCoy, Phys. Rev. A8 (1973)
  2526.}
\nref\rBazRes{V.V. Bazhanov and N.Yu. Reshitikhin, Int. J. Mod.
 Phys. A4 (1989) 115.}
\nref\rAcosetchar{E. Date, M. Jimbo A. Kuniba, T. Miwa and M. Okado,
 Nucl. Phys. B290 (1987) 231;\nl
 D. Kastor, E. Martinec and Z. Qiu, Phys. Lett. 200B (1988)
 434; \nl J. Bagger, D. Nemeshansky and S. Yankielowicz, Phys. Rev. Lett.
 60 (1988) 389.}
\nref\rGKO{P.~Goddard, A. Kent and D. Olive, Commun. Math. Phys. 103
 (1986) 105.}
\nref\rFatLyk{V.A. Fateev and S.L. Lykyanov, Sov. Sci. Rev. A Phys. 15
 (1990) 1.}
\nref\rGcosetchar{P.~Christe and F.~Ravanini, Int.~J.~Mod.~Phys.~A4
 (1989) 897.}
\nref\rBBSS{F.A.~Bais, P.~Bouwknegt, K.~Schoutens and M.~Surridge,
            Nucl.~Phys.~B304 (1988) 371.}
\nref\rBaxter{R.J. Baxter, {\it Exactly Solved Models in Statistical
 Mechanics} (Academic Press, London, 1982).}
\nref\rgepth{D. Gepner, Nucl. Phys. B296 (1988) 757.}
\nref\rKac{V.G. Kac and D.H. Peterson, Adv. in Math. 53 (1984) 125; \nl
 M. Jimbo and T. Miwa, Adv. Stud. in Pure Math. 4 (1984) 97; \nl
 V.G.~Kac and M.~Wakimoto, Adv.~in Math.~70 (1988) 156.}
\nref\rDV{R.~Dijkgraaf
 and E.~Verlinde, Nucl.~Phys.~B (Proc.~Suppl.) 5B (1988) 87.}
\nref\rISZ{C. Itzykson, H. Saleur and J.-B. Zuber, Europhys. Lett. 2
 (1986) 91.}
\nref\rRiSz{B.~Richmond and G.~Szekeres, J.~Austral.~Math.~Soc.~(Series~A)~31
             (1981) 362.}
\nref\rLewin{L. Lewin, {\it Dilogarithms and associated functions}
 (MacDonald, London, 1958).}
\nref\rBR{V.V.~Bazhanov and N.Yu.~Reshetikhin, J.~Phys.~A23 (1990) 1477
 and Progr. Theor. Phys. Suppl. 102 (1990) 301.}
\nref\rRav{F.~Ravanini, Phys. Lett. 282B (1992) 73.}
\nref\rZamtba{Al.B. Zamolodchikov, Nucl. Phys. B342 (1990) 695.}
\nref\rKlaMel{T.R.~Klassen and E.~Melzer, Nucl.~Phys.~B338 (1990) 485.}
\nref\rRSOStba{Al.B. Zamolodchikov, Nucl. Phys. B358 (1991) 497 and 524.}
\nref\rItoMox{H. Itoyama and P. Moxhay, Phys. Rev. Lett. 65 (1990) 2102.}
\nref\rcosetTBA{Al.B. Zamolodchikov, Nucl. Phys. B366 (1991) 122.}
\nref\rKR{A.N.~Kirillov and N.Yu.~Reshetikhin, J.~Phys.~A20 (1987) 1587.}
\nref\rKir{A.N.~Kirillov, Zap.~Nauch.~Semin.~LOMI 164 (1987) 121
 (J.~Sov.~Math.~47 (1989) 2450) and Cambridge preprint, hep-th/9212150.}
\nref\rFatZam{V.A. Fateev and Al.B. Zamolodchikov, Phys. Lett. 271B (1991) 91.}
\nref\rKlumPear{A.~Kl\"umper and P.A.~Pearce, J.~Stat.~Phys.~64
 (1991) 13; Physica A 183 (1992) 304.}
\nref\rFenInt{P. Fendley and K. Intriligator, Nucl. Phys. B372 (1992) 533.}
\nref\rKunNak{A. Kuniba and T. Nakanishi, Mod. Phys. Lett. A7 (1992) 3487.}
\nref\rFrenk{E. Frenkel and A. Szenes, Harvard preprint, hep-th/9212094.}
\nref\rAhn{C.~Ahn, Cornell preprint CLNS 91/1117, hep-th/9111022.}
\nref\rAhnNam{C.~Ahn and S. Nam, Phys. Lett. 271B (1991) 329.}
\nref\rChrMar{P.~Christe and M.J.~Martins, Mod.~Phys. Lett. A5 (1990) 157.}

\Title{\vbox{\baselineskip12pt\hbox{ITP-SB-93-05, RU-93-01}
\hbox{hep-th/9301046} } }
{\vbox{\centerline{Fermionic Sum Representations for}
\vskip 4mm
       \centerline{Conformal Field Theory Characters}}}

\vskip 6mm
\centerline{R.~Kedem,\foot{Institute for Theoretical Physics,
 SUNY, Stony Brook,  NY 11794-3840}~
 T.R.~Klassen,\foot{Department of Physics and Astronomy, Rutgers
 University, Piscataway, NJ 08854-0849} ~B.M.~McCoy,$^1$~ and ~E.~Melzer$^1$}

\vskip 17mm

\centerline{\bf Abstract}
\vskip 3mm
We present sum representations for all characters of the
unitary Virasoro minimal models. They can be viewed as fermionic
companions of the Rocha-Caridi sum representations, the latter
related to the (bosonic) Feigin-Fuchs-Felder construction.
We also give fermionic representations for certain characters of
the general ~${(G^{(1)})_k \times (G^{(1)})_l \o (G^{(1)})_{k+l}}$~
coset conformal field theories, the
non-unitary minimal models ${\cal M}(p,p+2)$ and ${\cal M}(p,kp+1)$,
the $N$=2 superconformal series, and the $\ZZ_N$-parafermion theories,
and relate the $q\to 1$ behaviour of all these fermionic sum
representations to the thermodynamic Bethe Ansatz.

\Date{\hfill 1/93}
\vfill\eject

\newsec{Introduction}

Recently it was found~\rKKMM\ that characters (or branching functions)
of the coset conformal field theories
{}~${(G^{(1)})_1 \times (G^{(1)})_1 \o (G^{(1)})_2}$, $G$ a
simply-laced Lie algebra,
can be represented in the form
\eqn\fsum{ {\sum_{\bf m}}^{Q}  ~ {q^{{1\o 2} {\bf m}B{\bf m}^t}
            \over{(q)_{m_1}\ldots (q)_{m_r}}} ~~,}
where $B$ is twice the inverse Cartan matrix of the algebra $G$, $r$ is the
rank of the algebra, ${\bf m}=(m_1,m_2,\dots ,m_r)$,
\eqn\qn{(q)_m~=~\prod_{k=1}^{m}(1-q^k)~,}
and $Q$ indicates   certain
restrictions on the summation variables ${\bf m}$ which
depend on the character under consideration. These expressions extend
the results of Lepowsky and Primc~\rLepPrim ~for the case $G=A_r$
to all simply-laced Lie algebras.
It was also shown in~\rKKMM~that these sums
can be interpreted as partition functions
of massless fermionic quasi-particles with non-trivial lower bounds on
the single-particle momenta, which depend on the number,
$m_a$, and type, $a$, of
quasi-particles present in a given state.  This interpretation was based on
an analysis~\rKedMc ~of Bethe's equations for the (gapless)
antiferromagnetic 3-state Potts chain whose continuum limit is the $c$=1
conformal field theory
of $\ZZ_4$-parafermions,
which corresponds to the case $G=A_3$ in the above notation.

These fermionic sum representations have widespread
applicability. The case of the Virasoro minimal models ${\cal M}(2,2r+3)$
was considered
in~\rFNO \rNRT, and the coset models ${(G^{(1)})_k\o U(1)^r}$
in~\rTerh \rKNS .
These results are all of the form~\fsum.
Here we generalize~\fsum\       to a form which encompasses a
much wider class of conformal field theories.
Specifically, we provide fermionic sum
representations for:

\medskip
\no $\bullet$ ~All of the characters of
the unitary Virasoro minimal models
${\cal M}(n+2,n+3)$, which correspond to the cosets
{}~${(A_1^{(1)})_n \times (A_1^{(1)})_1 \o (A_1^{(1)})_{n+1}}$;

\vskip 2mm
\no $\bullet$ ~Certain characters of ~${(G^{(1)})_k\times (G^{(1)})_l\o
(G^{(1)})_{k+l}}$~ for $G$ a simply-laced Lie algebra;

\no $\bullet$ ~Certain characters of non-unitary minimal
models ${\cal M}(p,p+2)$ and ${\cal M}(p,kp+1)$;

\no $\bullet$ ~The identity character in the unitary $N$=2 superconformal
series and in the $\ZZ_N$-parafermion theories.

\medskip

As with the previous work~\rKKMM~which led to \fsum, we are again motivated by
an analysis of Bethe's equations for the 3-state Potts chain,
this time for the ferromagnetic case~\rDKMM,
which is related through an orbifold construction~\rorbif ~to the
$r$=6 RSOS model at the III/IV boundary. The corresponding
partition function is the non-diagonal ($D$-series) modular
invariant combination~\rCar \rModInv ~of $c={4\o 5}$ Virasoro characters.

\bigskip
\newsec{Sum Representations for Virasoro Characters of ${\cal M}(n+2,n+3)$}

The characters $\chi^{(p,p')}_{~r,s}(q)$ of the
irreducible highest-weight representations
with $\Delta_{~r,s}^{(p,p')}={(rp'-sp)^2-(p'-p)^2\over {4pp'}}$
of the minimal model ${\cal M}(p,p')$
of central charge  $c_{p,p'}=1-{6(p-p')^2 \over {4pp'}}$
are~\rRoCa \rFFF
\eqn\Virchi{ \chi^{(p,p')}_{~r,s}(q) ~=~
   {q^{-\Delta_{~r,s}^{(p,p')}} \o (q)_\infty}    ~ \sum_{k\in \ZZ}
\left( q^{\Delta^{(p,p')}_{r+2kp,s}} -q^{\Delta^{(p,p')}_{r+2kp,-s}}\right)~~.}
The normalization is chosen such that the $\chi^{(p,p')}_{~r,s}(q)$
are power series in $q$ that start out with 1.
Whenever $p'$ in formulas below is suppressed,
it will be understood as being equal to $p+1$.

To present the fermionic representations for the characters \Virchi\
we recall the definition of the $q$-binomial
coefficient, that for integers $n$ and $m$
\eqn\qbin{ {n \atopwithdelims[] m}_q ~=~\cases{ ~~{(q)_n \o (q)_m (q)_{n-m}}
  ~~~~~~~~& if ~~$0\leq m \leq n$ \cr  ~~0 & otherwise,\cr} }
where $(q)_m$ is defined by \qn\ (with $(q)_0$=1).
We also set
${\infty \atopwithdelims[] m}_q=1/(q)_m$ for $m\geq 0$.
The $q$-binomial coefficients are polynomials in $q$, known as gaussian
polynomials~\randrews.

Our main result is that the characters \Virchi, for $p=p'-1=n+2$ with
$n$ a positive integer,
can all be written as special cases of the following general sum:
\eqn\Snauq{ S_n{{\bf Q}\atopwithdelims[]{\bf A}}({\bf u}|q) ~=~
 \sum_{{\bf m}\in (2\ZZ_{\geq 0})^n+{\bf Q}}
  q^{{1\o 4}{\bf m} C_n {\bf m}^t -{1\o 2}{\bf A}\cdot{\bf m}}
  ~{1\o (q)_{m_1}}~
  \prod_{a=2}^n ~
  { {1\o 2}({\bf m}I_n+{\bf u})_a \atopwithdelims[] m_a}_q~~,}
where
{}~${\bf A},{\bf u}\in \ZZ^n$,
{}~${\bf A}\cdot {\bf m}=\sum_{a=1}^n A_a m_a$,
{}~${\bf Q} \in (\ZZ_2)^n$ with
{}~$({\bf Q}I+{\bf u})_a\in 2\ZZ$~ for $a=2,\ldots,n$,
and ~$I_n$ and ~$C_n=2-I_n$ are the incidence and
Cartan matrix, respectively,
of the Lie algebra $A_n$.
Explicitly,
\eqn\Imat{ (I_n)_{ab} ~=~ \delta_{a,b+1} + \delta_{a,b-1} ~~~~
   {\rm for}~~ a,b=1,\ldots,n.}
Note that $S_n{{\bf Q}\atopwithdelims[]{\bf A}}({\bf u}|q)$ does
not depend on $u_1$; it will prove instructive to view
the factor $1/(q)_{m_1}$ in \Snauq\ as
{}~${(m_2+u_1)/2 \atopwithdelims[] m_1}_q$~ with $u_1$=$\infty$.

Due to the definition \qbin, the sum in \Snauq\ is actually restricted
to non-negative values of $m_a$, and furthermore for any given $m_1$
there is only a finite number of nonvanishing terms.
In fact, it is easy to see that the support
in the space of ${\bf m}$ of the summand in \Snauq\ is bounded by the region
defined by
$m_1\geq 0$ and $0\leq m_a \leq {n-a+1 \o n-a+2} ~m_{a-1}+(D{\bf u}^t)_a$
for $a=2,\ldots,n$, where $D_{ab}=\prod_{j=a}^b {n-j+1\o n-j+2}$~ for
$b\geq a$ and 0 otherwise.

To specify the full set of characters \Virchi\ denote
the $n$-dimensional
unit vector in the $a$-direction by ${\bf e}_a$
(\ie ~$({\bf e}_a)_b=\del_{ab}$), set ${\bf e}_a={\bf 0}$ for
$a \not\in \{1,\ldots,n\}$,
and let $\rhob={\bf e}_1+\ldots+{\bf e}_n$.
For fixed $n\geq 1$ define
\eqn\Qrs{ {\bf Q}_{r,s} ~=~ (s-1)\rhob+({\bf e}_{r-1} +{\bf e}_{r-3}+\ldots)
  + ({\bf e}_{n+3-s} +{\bf e}_{n+5-s}+\ldots)~~,}
where addition of the components of the vectors on the rhs is modulo
2. Our result is
that the
Virasoro characters \Virchi\
with $p-2=p'-3=n$
can be expressed as
\eqn\main{  \chi^{(p)}_{r,s}(q)~=~
  q^{-{1\o 4} (s-r)(s-r-1)}~
  S_n{{\bf Q}_{r,s}\atopwithdelims[]{\bf e}_{n+2-s}}
  ({\bf e}_r+{\bf e}_{n+2-s}|q)~~. }
(Note that terms in ${\bf u}$ which are proportional to ${\bf e}_1$
can be ignored.)
These representations have been conjectured on the basis of the
results of~\rDKMM ~and have been verified in many cases
to order $q^{100}$ or more using Mathematica.

Due to the symmetry $(r,s)$$\leftrightarrow$$(n+2-r,n+3-s)$ of the conformal
grid,
another representation must also exist, namely
\eqn\mainalt{             \chi^{(p)}_{r,s}(q)~=~
  q^{-{1\o 4} (s-r)(s-r-1)}~
  S_n{{\bf Q}_{s-1,r}\atopwithdelims[]{\bf e}_{s-1}}
  ({\bf e}_{s-1}+{\bf e}_{n+2-r}|q)~~, }
where we used ${\bf Q}_{n+2-r,n+3-s} = {\bf Q}_{s-1,r}$ (in $(\ZZ_2)^n$),
which follows from the definition \Qrs.

The equality of the rhs's of \main\ and \mainalt\ amounts to an
interesting set of identities between sums \Snauq\ with different
``characteristics''.
(For $(r,s)=(1,1)$ and $(r,s)=(n+1,1)$, labeling the corners of the
conformal grid, these identities are trivial).
The simplest example occurs already at $n$=1 for the character
of the spin field of the Ising model, where \main\ and \mainalt\ give
\eqn\chisig{ \chi_{1,2}^{(3)}(q) ~=~ \chi_{2,2}^{(3)}(q)
 ~=~ \sum_{{m=0\atop m~{\rm odd}}}^\infty {q^{m(m-1)/2} \o (q)_m}
 ~=~ \sum_{{m=1\atop m~{\rm even}}}^\infty {q^{m(m-1)/2} \o (q)_m} ~~.}
The equality of the two sums here is noted in the list of Slater~\rSlater ,
eqs.~(84)--(85). (It can be simply proved
using the methods of the next section, in particular
eq.~(3.12) with $M$=$\infty$.)

\newsec{Fermionic Quasi-Particle Interpretation }

The sum forms~\fsum\ have a
direct
interpretation
as the partition functions of fermionic quasi-particles~\rKKMM.
We now
present such   an
interpretation for the form \Snauq. Our
treatment follows the reverse of the procedure used in \rDKMM\ to
derive the characters for the special case $n=3$.

A quasi-particle form for an energy spectrum is a representation
of the
energy levels (above the ground state)
in the form
\eqn\En{E(\{P\})-E_{GS}
  ~=~\sum_{a=1}^n ~\sum_{{j_a=1 \atop {\rm rules}}}^{m_a} e_a(P_{a,j_a}) ~~}
with the total momentum given by
\eqn\Pn{P~=~\sum_{a=1}^{n}~\sum_{{j_a=1 \atop {\rm rules}}}^{m_a}P_{a,j_a} ~~,}
where the $P_{a,j_a}$ are chosen to satisfy a set of rules. When one of
the rules is
\eqn\fermi{P_{a,j_a}\neq P_{a,k_a}\quad {\rm for}\quad j_{a}\neq
k_{a}~,}
the spectrum is said to be fermionic.

The characters \main\ may be viewed
as partition functions constructed from the energies \En\  as
\eqn\pf{ Z(q)~=~\sum e^{-(E(\{P\})-E_{GS})/kT}~~.}
We will show that
\eqn\ZS{Z(q)= S_n{{\bf Q}\atopwithdelims[]{\bf A}}({\bf u}|q)~~}
if
\eqn\disp{ e_{a}(P) ~=~ vP ~~,}
where $v$ is the ``speed of sound (or light)'',
\eqn\qdef{ q=e^{-{2 \pi v\o LkT}}~,}
and the $P_{a,j_a}$
{}~($j_a=1,2,\ldots,m_a$      with      $m_a \equiv Q_a~({\rm mod}~2)$)
obey the exclusion principle \fermi\ but are otherwise
freely chosen from the sets
\eqn\Pone{P_{1,j_1} \in \Bigl\{ P_1^{\rm min}({\bf m}),
  ~P_1^{\rm min}({\bf m})+{2\pi \o L},
  ~P_1^{\rm min}({\bf m})+{4\pi \o L},~\ldots \Bigr\}~~,}
and for $2 \leq a \leq n$
\eqn\Pa{P_{a,j_a} \in \Bigl\{ P_a^{\rm min}({\bf m}),
  ~P_a^{\rm min}({\bf m})+{2\pi \o L},~\ldots,
  ~P_a^{\rm max}({\bf m}) \Bigr\}~.}
The vectors ${\bf P}^{\rm min,max}=\{P_a^{\rm min,max}\}$ here are
\eqn\Pmin{{\bf P}^{\rm min}({\bf m})
  ~ =~-{2\pi \over L}~{1\o 2} \Bigl({1\o 2}{\bf m}I_n+{\bf A}-\rhob \Bigr)}
where $\rhob$ denotes the $n$-dimensional vector $(1,1,\ldots,1)$,
and for $a \geq 2$
\eqn\Pmax{ P_a^{\rm max}({\bf m}) ~=~ -P_a^{\rm min}({\bf m})+
                  {2\pi \o L}({{\bf u}\o 2} -{ \bf A})_a~~.}

To derive this representation,
denote by
$Q_m(N;M)$  the number of partitions of the non-negative
integer $N$ into $m$ distinct
non-negative integers which   are
 smaller or equal to  the positive integer $M$, possibly infinite.
(We set $Q_0(N;M)=0$ for $N>0$, and ~$Q_0(0;M)=1$.)
The generating functions for $Q_m(N;M)$ are essentially
the $q$-binomial coefficients~\rStan :
\eqn\Qgen{ \sum_{N=0}^\infty Q_m(N;M) ~q^N ~=~ q^{{1\o 2} m(m-1)}~
  {M+1 \atopwithdelims[] m}_q~~.}
Then using \Qgen\ we can rewrite \Snauq\ as
\eqn\Snqp{ \eqalign{ S_n{{\bf Q}\atopwithdelims[]{\bf A}}({\bf u}|q)~ =
  \sum_{{\bf N}\in (\ZZ_{\geq 0})^n}~&\sum_{{\bf m}\in
   (2\ZZ_{\geq 0})^n+{{\bf Q}}}
  q^{ \rhob\cdot {\bf N}+{\bf m}\cdot
   {L\o 2\pi}{\bf P}^{{\rm min}}({\bf m})}~\cr
  \times~ & \prod_{a=1}^n ~ Q_{m_a}\bigl(N_a; {L\over 2\pi}
 (P_a^{{\rm max}}({\bf m})-P_a^{{\rm min}}({\bf m})) \bigr) ~~,} }
It is readily seen that the partition function $Z(q)$ of \pf,
constructed according to \disp--\Pa, is indeed equal
to the rhs of \Snqp.

{}From \Pone\ we see that there is one quasi-particle excitation with an
infrared cutoff at low momentum which depends on the number of
quasi-particles in the state. This is exactly the situation which
occurred in the computation of the branching functions~\rKKMM \rKedMc
{}~of ${(G^{(1)})_1\times (G^{(1)})_1\over {(G^{(1)})_2}}$,
where there are $r$=rank($G$) of these quasi-particles. The additional feature
of the present case is that the momenta $P_{a,j}$ for $2\leq a \leq n$
are restricted by \Pa\ to take on only a finite number of values
for given ${\bf m}$
(even
in the $L \rightarrow \infty$ limit). This phenomenon, while not
exactly what was originally envisioned in the term quasi-particle, is in
fact a common occurrence in quantum spin chains. It results from
cancellations seen in the earliest computations of energy levels in
the XYZ spin chain~\rJKM\ and is seen in all the RSOS models at the
boundary of the III/IV regimes~\rBazRes.
It also has a counterpart
in thermodynamic Bethe Ansatz analyses of factorizable scattering
theories (see sect.~5 for additional details).

Another feature not encountered in~\rKKMM \rKedMc~is that the momenta
$P_{a,j}$ can now take on negative values, even for $a$=1, and that the
`single-particle energies' are
$e_a(P)\! = \! v P$, not ~$e_a(P)\! =\! v |P|$.
A more
complete physical characterisation and interpretation of these
excitations is expected to be found from a
detailed study of the
energy levels of the 3-state chiral Potts model in which the spectrum can be
continuously  varied from a $\ZZ_4$ spectrum with 3 quasi-particles to the
above spectrum with one quasi-particle and two excitations with finite
momentum range.
Finally, we note that
these finite range excitations,
whose ${\cal O}(1/L)$ contribution to the energy was studied here,
also affect the degeneracy of the order
one energy levels. A more detailed discussion is found in~\rDKMM.

\newsec{Generalizations}

There are many generalizations of the results of
sect.~2. We will concentrate here mainly
on the character of the (extended) primary field creating the ground state
in various series of rational conformal field theories.
It appears that these characters can always be represented
as fermionic sums
\eqn\SB{ S_B({\bf u}|q) ~=~ \sum_{{\bf m}}
 ~S_B^{\bf m}({\bf u}|q)
  ~\equiv~ \sum_{{\bf m}}
  ~q^{{1\o 2} {\bf m} B {\bf m}^t} ~ \prod_{a=1}^n ~
  {{({\bf m}(1-B)+{{\bf u}\o 2})_a \atopwithdelims[] m_a}}_q ~~,}
where $B$ is a real  $n$ by $n$ symmetric matrix and ${\bf u}$ a vector whose
entries are either 0 or $\infty$. When
$B={1\o 2}C_n$ and ${\bf u}=(\infty,0,\ldots,0)$ the sum \SB\ is of
the form \Snauq, whereas taking
all $u_a$=$\infty$ it degenerates into the form of
\fsum.

We believe that all other characters can be obtained
by introducing nontrivial ``characteristics''
${\bf Q}, {\bf A}, {\bf u}$
as in \Snauq.
As an example, by elementary manipulations of the results
of~\rLepPrim~one can
show that up to an overall power of $q$
all the characters of the cosets
{}~${(A_r^{(1)})_1 \times (A_r^{(1)})_1 \o (A_r^{(1)})_{2}}$~
can be written in
the form \fsum\
with ${\bf m}$ in the quadratic form ${\bf m}C_r^{-1}{\bf m}^t$
replaced by
${\bf m}+{{\bf e}_a\o 2}$,
where ${\bf e}_a$ is an $r$-dimensional unit vector (or ${\bf 0}$).
Extending this observation to the coset
{}~${(E_8^{(1)})_1 \times (E_8^{(1)})_1 \o (E_8^{(1)})_2}$~ which is
identical to the minimal model ${\cal M}(3,4)$,
we obtain ~$\chi_{1,1}^{(3)}+\chi_{1,2}^{(3)}$~
if $m_1$ in ${\bf m}C_{E_8}^{-1}{\bf m}^t$
(in the basis used in~\rKKMM ) is replaced by
$m_1-{1\o 2}$,
and ~$\chi_{1,1}^{(3)}+\chi_{1,2}^{(3)}+\chi_{1,3}^{(3)}$~
if $m_2$ is replaced by $m_2-{1\o 2}$.
Thus, together with the result~\rKKMM~for $\chi_{1,1}^{(3)}$, we have
$E_8$-type fermionic representations for all three
characters of the Ising conformal field theory.
A more detailed analysis is left for future work.

\subsec{The cosets
  ~${(A_1^{(1)})_k \times (A_1^{(1)})_l \o (A_1^{(1)})_{k+l}}$.}

The characters of these coset models are given in~\rAcosetchar .
We conjecture that the identity character can be represented
in the form \SB\ with $B={1\o 2}C_{n}$, $n=k+l-1$, and $u_l$=$\infty$,
all other $u_a$ being 0.
For $l$=2 the series of theories labeled by $k$ is the
unitary $N$=1 superconformal
 series whose characters are given
in~\rGKO . We find that the character
corresponding to the identity superfield
in these models is obtained by summing over
$m_1\in \ZZ$, $m_a \in 2\ZZ$ ~for $a=2,\ldots,k+1$.

\subsec{The cosets
  ~${(G^{(1)})_k \times (G^{(1)})_l \o (G^{(1)})_{k+l}}$~
    with simply-laced $G$.}

The corresponding characters can be found
in~\rFatLyk \rGcosetchar~(in~\rFatLyk~the
coset models with $l$=1 are denoted by $[G_r^{(h+k)}]$ where $r$ is the
rank and $h$ the Coxeter number of $G$).
In this case we take
{}~$B=C_G^{-1} \otimes C_{n}$, $n=k+l-1$, where
$C_G$ is the Cartan matrix of $G$ and $C_n \equiv C_{A_n}$ as before.
Using a double-index notation
\eqn\BG{ B_{ab}^{\alp \beta} ~=~ (C_G^{-1})_{\alp \beta}
   ~(C_{n})_{ab} ~~~~~~~~~~\alp,\beta=1,\ldots,r={\rm rank}(G),~~~
    a,b=1,\ldots,n,}
we set ~$u^{\alp}_l$=$\infty$ ~for all
$\alp$ and 0 otherwise. (Cf.~\rTerh \rKNS ~where
{}~$B=C_G \otimes C_{n}^{-1}$~
with {\it all} $u_a^\alp$=0
is used
to produce fermionic sum representations
of the form \fsum\ for the identity characters in the cosets
${(G^{(1)})_{n+1} \o U(1)^r}$.)

As a non-trivial example consider the coset
{}~${(E_8^{(1)})_2 \times (E_8^{(1)})_1 \o (E_8^{(1)})_{3}}$~ of
central charge         $c= {21\o 22}$, which is
identified~\rBBSS ~as the Virasoro minimal model ${\cal M}(11,12)$ of
$(E_6,A_{10})$ type~\rModInv.
We verified to order $q^{100}$ that
 the corresponding sum \SB\ with all ~$m^\alp_a\in \ZZ$~
 agrees  with ~$\chi_{1,1}^{(11)}(q)+q^8 \chi_{1,7}^{(11)}(q)$, which
 is the extended identity character of this model.

\subsec{Non-unitary Virasoro Minimal Models ${\cal M}(p,p+2)$~ $(p$ odd $)$.}

The character
$\chi^{(p,p+2)}_{(p-1)/2,(p+1)/2}(q)$~  (see \Virchi) of the
field with lowest conformal dimension
{}~$\Delta^{(p,p+2)}_{(p-1)/2,(p+1)/2}=-{3\o 4p(p+2)}$~ in this model
is given by \SB\ with ~$B={1\o 2}C_{(p-1)/2}'$, where
$C_n'$ is the Cartan matrix
of the tadpole graph of $n$ nodes (\ie ~it differs from $C_n$ only
in one element, which is $(C_n')_{nn}$=1),
$u_1$=$\infty$ and $u_a$=0 for $a=2,\ldots,{p-1\o 2}$,
and
{}~$m_a \in 2\ZZ$.

Note that for the first model in this series, ${\cal M}(3,5)$, the sum
\SB\ reduces to the form of \fsum,
\eqn\chiBax{ \chi_{1,2}^{(3,5)}(q) ~=~ \sum_{m=0\atop m~{\rm even}}
  {q^{m^2/4} \o (q)_m} ~~.}
If the summation on the rhs of \chiBax\ is performed over odd instead
of even $m$, then one obtains
$q^{1/4}\chi_{1,3}^{(3,5)}(q)$.
Furthermore
\eqn\chiBaxt{ \chi_{1,1}^{(3,5)}(q) ~=~ \sum_{m=0\atop m~{\rm even}}
  {q^{(m^2+2m)/4} \o (q)_m} ~~,}
and  $q^{3/4}\chi_{1,4}^{(3,5)}(q)$
is obtained if one sums over
odd $m$ in \chiBaxt. The above four sums, representing all characters of
${\cal M}(3,5)$,
occur in~\rSlater\ and
have been encountered by Baxter~\rBaxter\ (see eq.~(14.5.50)) in regime IV
of the hard hexagon model.

\subsec{Minimal Models ${\cal M}(p,kp+1)$.}

For $k$=1 these are the unitary models considered in sect.~2. Here we
consider the cases $k=2,3,\ldots$ with $p>2$
(when $p$=2 the models are those treated in~\rFNO \rNRT ).
The character
$\chi^{(p,kp+1)}_{1,k}(q)$ of the field of lowest conformal
dimension in the model
is identified as \SB\ with $n=k+p-3$,
{}~$m_1,\ldots,m_{k-1}\in \ZZ$~ and ~$m_k,\ldots,m_{k+p-3}\in 2\ZZ$,
{}~$u_a$=$\infty$ for $a=1,\ldots,k$ and 0 otherwise. The nonzero
elements of $B$ are given by ~$B_{ab}=2(C_{k-1}'^{-1})_{ab}$ ~and
{}~$B_{ka}$=$B_{ak}$=$a$~ for ~$a,b=1,2,\ldots,k-1$, and
{}~$B_{ab}={1\o 2}[(C_{p-2})_{ab}+(k-1)\del_{ak}\del_{bk}]$~ for
{}~$a,b=k,k+1,\ldots,k+p-3$.

As in sect.~4.3, the case $p$=3 is special in that
the fermionic sums are of the form \fsum\ for any $k$.
We found that a slight modification of the matrix $B$ appropriate
for ${\cal M}(3,3k+1)$,
namely just setting $B_{kk}={k\o 2}$ while leaving all other elements
unchanged, gives the character $\chi_{1,k}^{(3,3k+2)}$ in ${\cal M}(3,3k+2)$.

\subsec{Unitary $N$=2 superconformal series.}

Expressions for the characters of these models,
of central charge ~$c={3k \o k+2}$~ where $k$ is a positive integer,
can be found in~\rgepth . The identity character,
given by ~$\chi_0^{0(0)}(q)+\chi_0^{0(2)}(q)$~ in the notation
of~\rgepth , can be obtained from \SB\ if one takes
{}~$B={1\o 2}C_{D_{k+2}}$, ~$u_k$=$\infty$ ~
and all other $u_a$ set to zero
(in the basis where the two nodes on the fork of the $D_{k+2}$ Dynkin
diagram are labeled by $k$+1 and $k$+2, and the junction node is labeled
by $k$),
and ~$m_{k+1},m_{k+2}\in \ZZ$~ while ~$m_a\in 2\ZZ$~ for all other $a$.

\subsec{$\ZZ_N$ parafermions.}

The characters of these models are the branching functions $b^l_m$~\rKac\
of the cosets
{}~${(A_{N-1}^{(1)})_1 \times (A_{N-1}^{(1)})_1 \o (A_{N-1}^{(1)})_{2}}$,
for which one type of fermionic sum representation is~\rLepPrim~of the
form \fsum.
The results of sect.~2 for the case $n$=3 provide an alternative
fermionic representation  for the characters of the
$\ZZ_3$-parafermion theory
which coincides with the minimal model ${\cal M}(5,6)$
with the $D$-series partition function.
(The $b^l_m$ in the case $N$=3 are linear combinations of the
$\chi_{r,s}^{(5)}$.)
This latter representation generalizes to arbitrary $N$; in particular,
the identity character
$b^0_0$ of the $\ZZ_N$-parafermions
is obtained from \SB\ by setting
{}~$B={1\o 2}C_{D_{N}}$, ~$u_{N}$=$\infty$ ~(in the basis used in sect.~4.5)
and all other $u_a$ set to zero, with $m_{N-1},m_{N}$ running
over all integers such that $m_{N-1}+m_{N}$ is even
while ~$m_a\in 2\ZZ$~ for $a=1,\ldots,N-2$.

\newsec{Behaviour as $q\to 1$ and Relation to Thermodynamic Bethe Ansatz}

Modular covariance relates the behaviour of a character as $q\to 1$ to
that of $q\to 0$, where the leading term is fixed by
the effective central charge~\rCar \rDV \rISZ
\eqn\ceff{\tilde{c} ~=~ c ~-~24 \Delta_{{\rm min}}}
 of the corresponding conformal field theory.
Here $c$ is the
central charge and
$\Delta_{{\rm min}}$ the lowest conformal dimension in the theory.
The $q\to 1$ behaviour of sum forms like~\fsum\ and \Snauq\ must be
consistent with this.
It was
noticed in~\rKKMM \rKedMc \rNRT~that the asymptotic analysis
of the sums of the form~\fsum\ considered there leads to the same
equations
for $\tilde{c}$ which had previously appeared in thermodynamic Bethe Ansatz
computations      of specific heats.
We here extend these observations to the sum forms \Snauq\ and \SB.
In fact, the results below inspired many of the generalizations in
sect.~4.

\medskip
Our analysis follows that of~\rNRT \rRiSz .
It is easy to see that the
leading behaviour of a sum like
$S_n{{\bf Q}\atopwithdelims[] {\bf A}}({\bf u}|q)$    in  \Snauq\
as $q\to 1$ is independent of ${\bf Q}$ and ${\bf A}$.
An asymptotic dependence on
${\bf u}$ exists only if some of its components are infinite.
Without loss of generality we can therefore consider sums
$S_B({\bf u}|q)$ of the form~\SB.

\medskip
Let $q=e^{2\pi i\tau}$ and
$\tilde{q}=e^{-2\pi i/\tau}$, with Im$\tau>0$. Then if
the coefficients in the series for $S_B({\bf u}|q)=\sum s_M q^M$
behave for large $M$ like $s_M \sim e^{2\pi\sqrt{\gamma M/6}}, ~\gamma>0$,
the series $S_B({\bf u}|q)$ diverges like
\eqn\SBlim{ S_B({\bf u}|q) ~\sim ~ \tilde{q}^{-\gamma/24}~~~~~~~~~
  {\rm as} ~~~q\to 1^-  ~. }
Here $\gam$ must equal the effective central charge \ceff\
of the corresponding conformal field theory.

The large $M$ behaviour of $s_M$ is found by considering
\eqn\sm{ s_{M-1}~=~\oint {dq \o 2\pi i} ~q^{-M} ~S_B({\bf u}|q) ~=~
  \sum_{{\bf m}\geq {\bf 0}} \oint {dq \o 2\pi i}~
                 q^{-M} ~S_B^{\bf m}({\bf u}|q)~~,}
where the contour
of integration is a small circle around 0. The behaviour of the integral
is now analyzed using a saddle point approximation. We first approximate
\eqn\lnSm{ \eqalign{ \ln  \Bigl(q^{-M} & S_B^{\bf m}({\bf u}| q)\Bigr)
  ~\simeq~ \bigl({1\o 2}{\bf m} B {\bf m}^t -M\bigr)\ln q \cr
  + & ~\sum_{a=1}^n
 \left( \int_0^{ ({\bf m}(1-B)+{{\bf u}\o 2})_a}-
        \int_0^{(-{\bf m}B+{{\bf u}\o 2})_a} -
        \int_0^{m_a} \right) dt ~\ln (1-q^t) \cr} }
for large ${\bf m}$, and set the derivatives of this expression
with respect to the $m_a$ to zero in order to find the saddle point.
Introducing ~$x_a={(1-w_a)v_a \o 1-v_a w_a}$~
and ~$y_a={1-w_a \o 1-v_a w_a}$~ where
{}~$v_a = q^{m_a}$~ and ~$w_a=q^{(-{\bf m}B+{{\bf u}\o 2})_a}$,
these extremum conditions reduce to
\eqn\xyeq{ 1-x_a ~=~ \prod_{b=1}^n x_b^{B_{ab}}~,~~~~~~~~
           1-y_a ~=~ \sigma_a \prod_{b=1}^n y_b^{B_{ab}}~~,}
where we define ~$\sigma_a$=0~ if ~$u_a$=$\infty$~ and 1 otherwise,
ensuring ~$y_a$=1 ~for ~$u_a$=$\infty$.

At the extremum point with respect to the $m_a$ we have
\eqn\lnSmext{ \eqalign{
 \ln \bigl( q^{-M} S_B^{\bf m}( &{\bf u}|q) \bigr) \Bigl|_{{\rm ext}} ~\simeq~
  -M\ln q \cr
 +~{1\o \ln q}\Biggl\{ & {1\o 2}\ln{\bf v} ~B~ \ln{\bf v}^t
 -\sum_{a=1}^n \bigl[{\cal L}(1-v_a)+{\cal L}(1-w_a)-{\cal L}(1-z_a)\bigr] \cr
 &-{1\o 2}\bigl[\ln{\bf v}\cdot\ln(1-{\bf v})+
               \ln{\bf w}\cdot\ln(1-{\bf w})-
               \ln{\bf z}\cdot\ln(1-{\bf z})\bigr] \Biggr\}~~ \cr} }
with ~$(\ln {\bf v})_a=\ln v_a$~ and
{}~$z_a = v_a w_a$, where
\eqn\dilog{ {\cal L}(z) = -{1\o 2} \int_0^z dt \left[
   {\ln t \o 1-t} + {\ln (1-t) \o t} \right]
  = -\int_0^z dt~{\ln (1-t)\o t}+{1\o 2}\ln z \ln(1-z)  }
is the Rogers dilogarithm function~\rLewin.
Now using \xyeq\ we see that the first term inside the braces in
\lnSmext\ cancels against the last.
Then using the five-term relation for the
dilogarithm~\rLewin\
\eqn\vterm{ {\cal L}(1-v)+{\cal L}(1-w)-{\cal L}(1-v w) ~=~
   {\cal L}(1-x)-{\cal L}(1-y)~~,}
where ~$x={(1-w)v \o 1-v w}$~ and ~$y={1-w \o 1-v w}$, we obtain
\eqn\lnSmexta{
 \ln\left(q^{-M} S_B^{\bf m}({\bf u}|q)\right)\Bigl|_{{\rm ext}} ~\simeq~
 -M \ln q - {\pi^2 \tilde{c} \o 6\ln q} ~~}
with
\eqn\ctilde{ \tilde{c} ~=~ {6\o \pi^2} \sum_{a=1}^n
    \left[ {\cal L}(1-x_a)-{\cal L}(1-y_a) \right]~~.}
Finally  the value of $q$ at the saddle point is determined by extremizing
\lnSmexta\ with respect to $q$,
which leads to ~$s_M \sim e^{2\pi\sqrt{\tilde{c}M/6}}$~ and
consequently to \SBlim\ with $\gamma=\tilde{c}$ of~\ctilde.

\medskip

It remains to check that \ctilde\ indeed gives the expected effective
central charge.
Consider first
the   general  coset conjecture of
sect.~4.2, where $B$ is given by \BG.
With
the $x_a$ written as $x_a^\alp$, where
{}~$a=1,\ldots,n\! =\! k\! +\! l\! -\!1$~ and
{}~$\alp \!=\! 1,\ldots,r\!=\!{\rm rank}(G)$,
eq.~\xyeq\ becomes
\eqn\genxeq{ 1-x_a^\alp ~=~ \prod_{\beta=1}^{r}~ \prod_{b=1}^{n}
               ~(x_b^\beta)^{(C_G^{-1})_{\alp\beta} (C_n)_{ab}} ~~.}
The $y_a^\alp$ satisfy the same equations as the $x_a^\alp$ except
for a factor $\sig_a^\alp$ on the rhs which enforces ~$y_l^\alp$=1.

The  system of equations
\ctilde --\genxeq\
has been encountered previously in the study
of the thermodynamics of RSOS spin chains~\rBR\ and
in the thermodynamic Bethe Ansatz of   integrable
perturbations of the
{}~${(G^{(1)})_k \times (G^{(1)})_l \o (G^{(1)})_{k+l}}$~
coset conformal field theories~\rRav\
(the particular case $k$=$l$=1 has been treated in~\rZamtba \rKlaMel ,
and the case of $G$=$A_1$ and arbitrary $k,l$
in [35-37]).
In the latter framework these equations arise in an analysis of
the corresponding
factorizable scattering theories
where there are $r$ particles which
are associated with the labels $(\alp,a$=$l$) in \genxeq.
All other labels, corresponding to the excitations of finite momentum
range in the context of
sect.~3, can be thought of as associated with
fictitious particles which are sometimes referred to as
``pseudoparticles''~\rRSOStba ~or ``magnons''~\rRav .

Finally, in order to reproduce \ceff\ the sums of dilogarithms must be
evaluated. This subject has been extensively investigated
[29-31][34][38-44].
Here we use the sum rule~\rBR \rKir
\eqn\cGn{ c(G,n) ~\equiv~ {6\o \pi^2}\sum_{\alp=1}^{r}~ \sum_{a=1}^n
 {\cal L}(1-x_a^\alp) ~=~ (n+1) r ~-~ c(G^{(1)}_{n+1}) ~}
where
\eqn\cGk{c(G^{(1)}_k) ~=~ {k ~{\rm dim}(G)\o k+h}}
is the central charge of the level~$k$
 WZW model based on $G$,
dim($G$) is the dimension of $G$ and $h$ its (dual) Coxeter number.
As in~\rBR \rRav\ we therefore find
\eqn\genc{\tilde{c} ~=~ c(G,n)-c(G,k-1)-c(G,l-1) ~=~
             c(G^{(1)}_{k}) + c(G^{(1)}_{l}) - c(G^{(1)}_{k+l})~,}
\ie~the central charge of the
{}~${(G^{(1)})_k\times (G^{(1)})_{l}\o (G^{(1)})_{k+l}}$~ coset conformal field
theory.
(For these unitary coset models $\Delta_{\rm min}$=0 and
so $\tilde{c}=c$.)
Explicitly for $G=A_r$, for instance, \genc\ reads
\eqn\ctAr{ \tilde{c}~=~r(r+2) \Bigl( {k\o k+r+1}+{l\o l+r+1}
   -{k+l\o k+l+r+1} \Bigr)~~.}

\medskip
For the (non-unitary) Virasoro minimal models
${\cal M}(p,p')$ the effective central charge is
{}~$\tilde{c}=1-{6\o pp'}$. For the particular series
of sects.~4.3 and~4.4, the
dilogarithm sum rules which
reproduce the correct~$\tilde{c}$ ~for
${\cal M}(p,p+2)$
and ${\cal M}(p,kp+1)$ have been encountered in the thermodynamic
Bethe Ansatz calculations of refs.~\rAhn ~and~\rAhnNam , respectively
(for $k$ odd in the latter case, but the sum rule holds for $k$ even as well).
The case ${\cal M}(3,5)$ has also been treated in~\rChrMar , and
we checked that our choice of the matrix $B$ appropriate for ${\cal M}(3,3k+2)$
(cf.~sect.~5.4)
leads to the expected effective central charge also for $k>1$.
Finally, the dilogarithm sum rules which are relevant for the
theories discussed in sects.~4.5 and 4.6 can be found
in~\rFenInt~and~\rFatZam , respectively.

\bigskip
There are numerous other examples of systems of equations of the form
\xyeq,\ctilde\ which arise in thermodynamic Bethe Ansatz
computations, to which
the general method of this section can be applied
to obtain (conjectures for) fermionic sum  representations of
characters.

\vskip 20mm

{{\vbox{\centerline{\bf Acknowledgements}}}
\nobreak\medskip
We are pleased to acknowledge fruitful discussions with Dr.~G.~Albertini
and Dr.~S.~Dasmahapatra,
and thank Prof.~A.~Kuniba for a useful comment.
The work of R.K.~and B.M.M.~was partially supported by the
National Science Foundation under grant DMR-9106648.
The work of T.R.K.~was supported by NSERC and the Department of Energy,
grant DE-FG05-90ER40559, and
that of E.M.~by NSF grant 91-08054.}

\bigskip

\vfill
\eject
\listrefs

\vfill\eject

\bye\end